\documentclass[aps,prl,10pt,nofootinbib,
twocolumn,unsortedaddress]{revtex4-1}

\usepackage{epsfig}
\usepackage{bm}
\usepackage{latexsym}
\usepackage{natbib}
\usepackage{url}
\usepackage{dcolumn}
\usepackage{color}
\usepackage[usenames,dvipsnames]{xcolor}
\usepackage{amsfonts,amssymb,amsmath}
\usepackage{graphicx,epsfig}
\usepackage{psfrag}
\usepackage{subfigure}
\usepackage{comment}
\usepackage[utf8]{inputenc}
\usepackage{amsmath}
\usepackage{hyperref}
\hypersetup{colorlinks=true}

\def\be{\begin{equation}}
\def\ee{\end{equation}}
\def\gev{{\rm \,Ge\kern-0.125em V}}

\newcommand{\gsim}{\mbox{\raisebox{-.6ex}{~$\stackrel{>}{\sim}$~}}}

\begin{document}
 
\title{A note on Single-field Inflation and the Swampland Criteria}

\author{Suratna Das}%
 \email{suratna@iitk.ac.in}
\affiliation{Department of Physics, Indian Institute of Technology, Kanpur 208016, India}

\date{\today}

\begin{abstract}
The recently proposed Swampland Criteria aim to evade any (meta-)stable de Sitter constructions within String landscapes, making it difficult to accommodate accelerating phases, like dark energy domination and inflationary epoch, in cosmology. In this note, we analyse the status of various models of single field slow-roll inflation given the old as well as the refined Swampland conjectures, which constrain the form of scalar potentials in any low energy effective field theory residing in the landscapes.  In particular, we note that Warm Inflation turns out to be the most befitting scenario as long as lifting the tensions with Swampland Criteria are concerned. 
\end{abstract}

\maketitle


Despite the remarkable agreement with the current data, it seems that the cosmological inflationary paradigm is yet to pass a few litmus tests, some of which are age-old issues\footnote{Such as the initial condition problem \cite{Ijjas:2013vea, Brandenberger:2016uzh, Das:2018kyl} and the quantum-to-classical transition of primordial perturbations \cite{Martin:2012pea, Das:2013qwa}} and some of them have been coined very recently, dubbed the Swampland Criteria. These Swampland Criteria, initially proposed in \cite{Obied:2018sgi} and later a refined version in \cite{Ooguri:2018wrx}, caused enough ruckus in the scientific community, as these criteria, if proved to be true, can potentially jeopardise our present understanding of the universe, both present as well as early (exponential expansions), as has been claimed in \cite{Agrawal:2018own}. 

String Theory, probably the best known theory of quantum gravity to date, provides us with a vast `{\it landscape}', with nearly $10^{500}$ vacua, where consistent quantum theory of gravity is believed to be formulated with consistent low-energy Effective Field Theories (EFTs). But such `{\it landscapes}' are known to be surrounded by even larger regions, dubbed `{\it swamplands}', where apparently consistent EFTs, which are coupled to gravity, are, in reality, inconsistent with quantum theory of gravity. Hence, it is desirable for consistent EFTs not to lie in the `swamplands', which eventually has led us to some set of conjectures, such as the weak gravity conjecture proposed a decade ago \cite{ArkaniHamed:2006dz}, and another set of `Swampland Criteria' which has been proposed quite recently \cite{Obied:2018sgi, Ooguri:2018wrx}. It has been noted that though the Minkowski and anti-de Sitter vacua can be easily obtained in String landscapes, it is a taxing job to embed de Sitter vacua in these coveted regions \cite{Brennan:2017rbf}. Such observations has led to the belief that de Sitter vacua rather reside in the `{\it swamplands}', which makes the case difficult for the theories or paradigms relying on de Sitter solutions. 

The simplest class of inflationary paradigm, known as the single-field slow-roll models, depicts an early phase of (nearly) de Sitter evolution of the universe, governed by a dynamical scalar field (known as the inflaton field) slowly-rolling down its nearly-flat potential. It is desirable to put the inflationary models in a UV complete field theory, where the low-energy EFT leads to inflation. It is thus obvious that not being able to obey the above mentioned set of conjectures, dubbed the Swampland Criteria, instantly poses threats to inflationary paradigm. Conversely, not being able to accommodate observationally supported inflationary paradigm within String landscapes can also pose threat to String Theory constructions. Hence, it is an important task to recheck the status of inflationary paradigm given the String Swampland Criteria  \cite{Obied:2018sgi, Ooguri:2018wrx}.

The first set of Swampland Criteria was proposed in \cite{Obied:2018sgi} which were as follows :
\begin{itemize}
\item {\bf Swampland Criterion I (The Distance Conjecture)}: This criterion limits the range traversed by scalar fields in field space as 
\begin{eqnarray}
\frac{\left|\Delta\phi\right|}{M_{\rm Pl}}\lesssim \Delta.
\label{sc1}
\end{eqnarray}
\item {\bf Swampland Criterion II (The de Sitter Conjecture)}: This criterion limits the gradient of scalar potentials in an EFT as 
\begin{eqnarray}
M_{\rm Pl}\frac{\left|V_\phi\right|}{V}\gsim c.
\label{sc2}
\end{eqnarray}
\end{itemize}
where $M_{\rm Pl}$ is the reduced Planck mass, and $V_\phi\equiv dV/d\phi$. Here $\Delta,\, c\sim {\mathcal O}(1)$. The actual value of $c$, which is a constant, depends on the details of compactification and often turns out to be greater than $\sqrt{2}$ \cite{Obied:2018sgi}. It can be readily seen why the second Swampland Criterion is of instant threat to inflationary paradigm. The slow-roll parameter $\epsilon$ of the inflaton field, which is a measure of the slop of the potential, is defined as 
\begin{eqnarray}
\epsilon=\frac12 M_{\rm Pl}^2\left(\frac{V_\phi}{V}\right)^2,
\end{eqnarray}
and $\epsilon<1$ ensures the slow-roll of the inflaton field during inflation. Thus the slow-roll condition is in direct conflict with Swampland Criterion II. 

However, demanding that no critical de Sitter vacuum exists, allows one to consider values of $c$ smaller than unity, until it is positive. Thus one can even consider $c\sim\mathcal{O}(10^{-1})$, which is as good as $c\sim {\mathcal O}(1)$ \cite{Kehagias:2018uem}. Even this seems to be not in accordance with the current observation of tensor-to-scalar ratio ($r$). Single-field slow-roll models with canonical kinetic term gives rise to both tensor and scalar perturbations with a ratio of their amplitudes as 
\begin{eqnarray}
r=16\epsilon.
\label{r-eps}
\end{eqnarray}
This yields 
\begin{eqnarray}
r=16\epsilon\gsim 8c^2\sim 0.08,
\end{eqnarray}
considering $c\sim10^{-1}$. 
Observations by PLANCK satellite and BICEP2/KEK Array ground-based small aperture telescopes together put an upper bound on this tensor-to-scalar ratio $r\lesssim 0.064$ \cite{Akrami:2018odb}. 
Thus apparently the Swampland Criterion II is in direct conflict with the current data as has been pointed out in \cite{Agrawal:2018own}. The lower the upperbound on $r$ to be set by the future observations, the more the tension with the criterion would become. This realisation has led to a flurry of papers where several single-field models have been analysed just to conclude that they are indeed inconsistent with the Swampland Criterion II \cite{Agrawal:2018own, Achucarro:2018vey, Garg:2018reu, Dias:2018ngv, Kehagias:2018uem, Kinney:2018nny}. \footnote{It is to note that the Swampland Criterion I is not in direct tension with the observations, as writing $\Delta\phi\sim 30\sqrt{r/2}\, M_{\rm Pl}$ \cite{Achucarro:2018vey, Kehagias:2018uem}, we see that the factor $30\sqrt{r/2}$ is $\mathcal{O}(1)$.} 

However, it was also readily observed that by going beyond the single-field domain, the multifield models of inflation can be accommodated with the swampland criterion in question. In \cite{Kehagias:2018uem}, the authors considered {\it curvaton models} \cite{Lyth:2001nq, Lyth:2002my}, wherein the primordial perturbations are generated by a different scalar field, called the {\it curvaton}, at the end of inflation when the curvaton isocurvature perturbations get converted into curvature perturbations, and the single-field consistency relation, given in Eq.~(\ref{r-eps}), does not hold in such models \cite{Byrnes:2014xua}. 
Hence, such models are not constrained by the relation given in Eq.~(\ref{r-eps}) and the observation of $r$ is irrelevant to the validity of the Swampland Criterion II. It is observed in \cite{Achucarro:2018vey} that in the multifield model of inflation the relation between $r$ and $\epsilon$ turns out to be 
\begin{eqnarray}
r=16\epsilon c_s,
\label{r-multi}
\end{eqnarray}
where $c_s$ is the speed of sound at which the curvature perturbations propagate in a multifield model. This $c_s$ can be written as 
\begin{eqnarray}
c_s=\left(1+\frac{4\Omega^2}{M_{\rm Pl}^2}\right)^{-1/2},
\end{eqnarray}
where $\Omega$ is the rate of turning of the inflationary trajectory in the multifield space. As $c_s$ is less than unity (or at best 1), this gives $r<16\epsilon$ ($r=16\epsilon$), thus lifting the tension between the criterion given in Eq.~(\ref{sc2}) and the observational upper bound on $r$ for cases with $c_s<0.8$.

Hence certainly, one way to tackle the Swampland Criterion II is by going beyond the simplest single-filed models. And yet, as the simplest single-field models are the ones most favoured by the data it calls for seeking single-field scenarios which would be in accordance with Swampland Criterion II. Recently it is proposed in \cite{Brahma:2018hrd} that non-Bunch-Davies initial state for cosmological perturbation yields \cite{Ashoorioon:2013eia, Ashoorioon:2014nta}
\begin{eqnarray}
r=16\epsilon\gamma,
\label{r-nbd}
\end{eqnarray}
where the factor $\gamma$ turns out to be the ratio of the Bogoliubov transformations of the tensor and scalar perturbations:
\begin{eqnarray}
\gamma=\frac{\left|\alpha_k^{(t)}+\beta_k^{(t)}\right|^2}{\left|\alpha_k^{(s)}+\beta_k^{(s)}\right|^2},
\end{eqnarray}
where $\gamma$ can be made less than unity by choosing some proper inflationary model. Thus such models featuring non-Bunch-Davies vacuum can make the single-field scenario consistent with Swampland Criterion II. This claim has been counteracted in a recent analysis \cite{Ashoorioon:2018sqb} where it was shown that such non-Bunch-Davies vacuum for scalar perturbations would yield large local non-Gaussianities, making them incompatible with present observations. On the other hand, considering non-Bunch-Davies vacuum only for tensor modes can make such models to be in tune with both Swampland Criteria and current bounds on primordial non-Gaussianities, as that would generate large flattened non-Gaussianities only in the tensor sector \cite{Ashoorioon:2018sqb}. 

The case of eternal inflation \cite{Vilenkin:1983xq, Guth:2007ng} has also been discussed in literature \cite{Matsui:2018bsy, Dimopoulos:2018upl} as a case of single field inflation in the context of Swampland Criteria. It is well known that eternal inflation, where inflaton quantum fluctuations dominate over the classical dynamics of the inflaton field, cannot yield the observed primordial scalar spectrum as it generates way too large scalar amplitude ($\mathcal{O}(1)$, where  the observed spectrum is $\sim \mathcal {O}(10^{-9})$), and thus should be followed by a phase of standard slow-roll inflation. Even though, eternal inflation helps explain the initial condition for slow-roll inflation and also helps populating the large number of String landscape vacua, one of which corresponds to our own universe. As the Swampland Criteria are universal within a given EFT, it is thus an important task to investigate whether eternal inflation can be realised within String landscapes, though it is not bounded by the cosmological observations. As has been discussed in both these works \cite{Matsui:2018bsy, Dimopoulos:2018upl}, eternal inflation can only be accommodated in landscapes if $c\lesssim\mathcal{O}(10^{-2})$, demanding that during eternal chaotic inflation the potential should be below Planck scale \cite{Matsui:2018bsy} and demanding a quantum jump timescale to be less than the inverse of the Hubble parameter during eternal inflation in a steep potential \cite{Dimopoulos:2018upl}.

The purpose of this note is to point out couple of other single-field scenarios which accommodate the Swampland Criterion II nicely.
One can easily note from Eq.~(\ref{r-multi}) and Eq.~(\ref{r-nbd}) that it is the suppression factor $c_s$ or $\gamma$ which is playing the trick to make either a multi- or a single-field model compatible with the swampland criterion. Hence, one can seek for similar suppression factors arising in single-field scenarios which can make the respective models compatible with the Swampland Criteria II. The two such scenarios are:
\begin{itemize}
\item {\bf $k-$inflation}: The obvious choice, in such a case, would be the $k-$inflation scenario \cite{Garriga:1999vw} where the curvature perturbations travel with subluminal speed of sound due to non-canonical kinetic terms of the inflaton field, yielding  
\begin{eqnarray}
r=16\epsilon c_s
\end{eqnarray}
again, with $c_s<1$. This has also been noted in \cite{Kinney:2018nny} while discussing the case of DBI (Dirac-Born-Infeld) Inflation. 
\item {\bf Warm Inflation}: Warm Inflation \cite{Berera:1995ie} scenario is even more interesting as far as handling the Swampland Criterion II is concerned, as this scenario is capable of tackling the situation in three different ways, as we state below. 

Warm inflation is an alternative scenario to the generic cold inflation scenario, where the inflaton field dissipates to a thermal bath while inflating and thus maintains a constant radiation energy density throughout inflation despite the exponential expansion. Thus such a scenario does not require to call for a reheating phase at the end of inflation. In standard cold inflation scenario a inflaton field is in need of a minimum of the potential where the field would roll down at the end of inflation to reheat the universe by oscillating at the bottom of the potential and dissipating its energy to other fields. According to Swampland Criterion II, scalar fields should not have any such minima in their potentials. As Warm Inflation does not call for a reheating phase, it naturally is not in need of any such minima of the inflaton potential, which is rather essential in a cold inflationary set up.

Without going into the details of model building of Warm Inflation scenario, one can simply write down the dissipative equation of motion of the inflaton field as 
\begin{eqnarray}
\ddot\phi+3H\dot\phi+\Gamma\dot\phi+V_\phi=0,
\end{eqnarray}
where $\Gamma$ is the dissipative co-efficient which amounts to an extra friction term in the equation of the inflaton field. Defining 
\begin{eqnarray}
Q=\frac{\Gamma}{3H},
\end{eqnarray}
the slow-roll of the inflaton field is achieved when the following criteria are satisfied \cite{Bartrum:2013oka}:
\begin{eqnarray}
\epsilon&=&\frac{M_{\rm Pl}^2}{2}\left(\frac{V_\phi}{V}\right)^2\ll1+Q,\nonumber\\
|\eta|&=&M_{\rm Pl}^2\left(\frac{|V_{\phi\phi}|}{V}\right)\ll1+Q,\nonumber\\
\sigma&=&M_{\rm Pl}^2\left(\frac{V_\phi}{\phi V}\right)\ll 1+Q,\nonumber\\
\beta&=&M_{\rm Pl}^2\left(\frac{\Gamma_\phi V_\phi}{\Gamma V}\right)\ll1+Q.
\label{sr-warm}
\end{eqnarray}
Thus it can be easily seen that $\epsilon<1$ is no longer the criteria to be met to yield the slow-roll of the inflaton field. 
Above all, in the strong dissipative regime, when $\Gamma>3H$ and $Q>1$, then even with $c\sim \mathcal{O}(1)$ the Swampland Criterion II would not meddle with the slow-rolling of the inflaton field. This is a definite advantage of Warm Inflation over all the other scenarios discussed so far in the context of Swampland Criteria. 

Besides, the Warm Inflation scenario yields a tensor-to-scalar ratio as \cite{Bartrum:2013oka}
\begin{eqnarray}
r=\left(\frac{H}{T}\right)\frac{16\epsilon}{(1+Q)^{5/2}},
\end{eqnarray}
where $T$ is the temperature of the thermal bath with $T>H$ (and $Q$ is, of course, positive). Hence, it can be easily seen that one of the features of Warm Inflation is $r<16\epsilon$, which is in favour of the Swampland Criterion II as far as the cosmological observations are concerned. 

As a passing comment, Warm Inflationary scenario can also be successfully realised in a large class of String Theory models \cite{BasteroGil:2009gh, BasteroGil:2011mr, Cai:2010wt}.
\footnote{After this analysis was presented, which points out several advantages of warm inflation over the standard cold inflation given the Swampland Criteria, two more analysis of warm inflation in the light of Swampland have been done, one by Motaharfar et al. \cite{Motaharfar:2018zyb} and the other by the author of this manuscript \cite{Das:2018rpg} analyzing the parameter space of warm inflation to best fit the Swampland Criteria.  Motaharfar et al. \cite{Motaharfar:2018zyb}, without relating with the observational upper bound on $r$, concludes that taking both the Swampland Criteria together would drive warm inflation to take place deep into the strong dissipative regime where $Q$ should be larger, at least of the order of the minimum number of efolds required. Such strong dissipative regime warm inflation scenarios can yield scale depend scalar power spectrum which would contradict with the observations. On the other hand, the analysis made in \cite{Das:2018rpg}, shows that treating the two Swampland Criteria separately with the observational upper bound on $r$ actually requires $1+Q$ to be slightly greater than unity. Such weak dissipative regime $(Q\leq1)$ of warm inflation is more in accordance with current observations.}
\end{itemize}

{\it After the refinement:} It was later pointed out in \cite{Garg:2018reu}, that in order to constrain de Sitter/inflating vacua in String Theory, which has given rise to the two Swampland Criteria we were discussing so far, one should rather restrict `slow-roll' altogether. The Swampland Criterion II, or the de Sitter conjecture, does indeed attempt to do that by demanding $\epsilon>1$. But, slow-roll depends on two slow-roll parameters $\epsilon$ and $\eta$ $(\equiv M_{\rm Pl}^2V_{\phi\phi}/V)$, where one requires $\epsilon\ll1$ and $|\eta|\ll1$ for slow-roll. Hence in order to restrict `slow-roll' either of the two slow-roll parameters should be greater than one. 

Followed by this argument of refining the Swampland Criterion II based on slow-roll arguments \cite{Garg:2018reu}, a refined version of de Sitter conjecture has now been proposed in \cite{Ooguri:2018wrx}, where it was shown that the refined Swampland Criterion II directly follows from implementing Swampland Criterion I or the distance conjecture which has been more firmly established in many String Theory constructions. The distance conjecture suggests that scalar fields (moduli) travelling to large (trans-Planckian) geodesic distances give rise to tower of light states with masses $m\sim e^{-a\Delta\phi}$, where $\Delta\phi$ is the change in the field value in Planck units and $a\sim\mathcal{O}(1)$. If the accelerating universe has causal region with an apparent horizon of radius $R$, then the number of effective degrees of freedom, increased by having these towers of light particles with exponentially small masses, increases the entropy within the causally connected region. This increased entropy then influences how the scalar potential behaves in any weak coupling limit. In order to see the effect on the potential, the effective number of particle species (below the cut-off of the effective theory) and entropy coming from the towers of particles were parametrised as \cite{Ooguri:2018wrx}
\begin{eqnarray}
N(\phi)&=&n(\phi)e^{b\phi},\nonumber\\
S_{\rm tower}(N,R)&\sim& N^{\gamma}R^\delta,
\end{eqnarray}
respectively, where $n(\phi)$ is the effective number of towers of states that are becoming light and $b$ depends on the mass gaps and other features of the towers and is often different from $a$. However, the Bousso bound \cite{Bousso:1999xy} suggests that $S_{\rm tower}\leq R^2$ and the distance conjecture suggests that $n(\phi)$ should increase monotonically as $\phi$ increases. Thus demanding both $S_{\rm tower}\leq R^2$ and $dn(\phi)/d\phi>0$ and knowing that for de Sitter horizon $R^2\sim H^{-2}\sim V^{-1}$, one arrives at the condition 
\begin{eqnarray}
M_{\rm Pl}\frac{|V_\phi|}{V}>\frac{2b\gamma}{\delta-2}\equiv c,
\end{eqnarray}
which is the old de Sitter conjecture or Swampland Criterion II and violates the slow-roll condition $\epsilon<1$. On the other hand, to have a stable accelerating vacua, the semi classical picture should not break down due to large quantum corrections of $\phi$. It can be seen from the equation of motion of the quantum modes of the scalar field that the quantum modes become tachyonic on horizon crossing if $V_{\phi\phi}$ becomes lesser than $-c'H^2\sim-c'/R^2\sim -c'V$. Thus a stable semi-classical picture would call for $M_{\rm Pl}^2V_{\phi\phi}\geq-c'V$ with $c'\sim\mathcal{O}(1)$, which implies that the condition $V_{\phi\phi}\leq-c'V$ would evade any (meta-)stable de Sitter vacua. 
Thus the refined de Sitter conjecture is now read as 
\begin{eqnarray}
M_{\rm Pl}\frac{|V_\phi|}{V}>c \quad{\rm or} \quad M_{\rm Pl}^2\frac{{\rm min}(V_{\phi_i\phi_j})}{V}\leq-c',
\end{eqnarray}
where both $c$ and $c'$ to be of the order unity, which also violates the slow-roll conditions. Thus the refined de Sitter conjecture restricts the form of the scalar potential within a causal region (in a weak coupling regime). 

The above conditions can be written in terms of slow-roll parameters within the framework of single-field inflation as 
\begin{eqnarray}
\epsilon>\frac{c^2}{2} \quad{\rm or} \quad \eta \leq-c'.
\label{refined}
\end{eqnarray}
We had been discussing the first condition, i.e. $\epsilon>c^2/2$ (in which case $\eta$ can be greater than $-c'$), so far, and have noticed that it is quite difficult to accommodate single field inflation in String landscapes given this criteria, except for few cases like non-Bunch-Davies vacuum for tensor modes, $k-$inflation and warm inflation. It is to note that though the refined de Sitter conjecture puts bound on the form of the potential and hence restricts the slow-roll dynamics, the arguments, as presented in \cite{Ooguri:2018wrx} and as has been illustrated above, depends upon the size of the causal horizon $R$ and the entropy inside it which is restricted by the Bousso bound  \cite{Bousso:1999xy}. For $k-$inflation the scalar modes travel with a speed $c_s<1$, and thus has a `sound horizon' ($c_sH^{-1}$) within which they are causally connected \cite{Garriga:1999vw}. Thus apparently the refined de Sitter conjectures as well as the Bousso bound (which depends on the size of the causal region) might take a different form in such scenarios, and we defer this analysis for a future study.

We now turn towards the second criteria $ \eta \leq-c'$ with $\epsilon\ll1$, which helps evade the contradiction coming from the observed upper-bound on $r$ (which demands $\epsilon<0.004$). First of all, 
this option implies that the inflaton potential must be concave. But, $\eta$ is directly related to another primordial observable, namely the scalar spectral index, which for single field model turns out to be 
\begin{eqnarray}
n_s-1=2\eta-6\epsilon,
\end{eqnarray}
and for $\epsilon\ll1$, this would be $n_s-1\approx2\eta$. The scalar spectral tilt is observationally a well constrained primordial parameter and according to the recent Planck observations $n_s=0.9649\pm0.0042$ at 68$\%$ CL \cite{Akrami:2018odb}. Hence, demanding that $|\eta|\geq\mathcal{O}(1)$, to be in tune with the refined de Sitter Criterion, one faces direct contradiction with current measurement of the scalar spectral tilt \cite{Fukuda:2018haz, Agrawal:2018rcg, Chiang:2018lqx}. This option of the refined de Sitter conjecture thus turns out to be even more difficult to accommodate than the previously stated de Sitter conjecture ($\epsilon>c^2/2$) as far as single field inflation is concerned. It was even emphasised in \cite{Chiang:2018lqx} that, single field inflation cannot be accommodated in the String landscapes, given the refined Swampland criteria, without fine-tuning one of the parameters $c$ and $c'$, both of which the refined criteria claims to be of order unity. Thus, even treating both these conditions, appearing in refined de Sitter conjecture (Eq.~(\ref{refined})), separately, one runs into contradiction with observations as well as with the criteria itself, unless one fine-tunes the parameters $c$ or $c'$ to at least two orders lower than unity. Other single field scenarios studied in this context, like Higgs inflation \cite{Cheong:2018udx}, Type I Hilltop Inflation \cite{Lin:2018rnx} and Minimal Gauge Inflation \cite{Park:2018fuj}, also suggest fine-tuning of both or either of the parameters to accommodate the respective scenarios within String landscapes.

The option $ \eta \leq-c'$ with $\epsilon\ll1$ also makes curvaton model to run into a conflict with observations as for such models one has 
\begin{eqnarray}
n_s-1=2\frac{\dot H}{H^2}+2\eta_{\sigma\sigma},
\end{eqnarray}
where $\sigma$ is the curvaton field, and with $\epsilon\sim-\dot H/H^2\ll1$, one gets $n_s-1\approx 2\eta_{\sigma\sigma}$ which again to meet the de Sitter conjecture should of the order unity \cite{Fukuda:2018haz}. Thus curvaton models perform better under the condition $\epsilon>c^2/2$. However, a recent analysis \cite{Kinney:2018kew} shows that this option, $ \eta \leq-c'$ with $\epsilon\ll1$, can make hilltop eternal inflation in accordance with the refined Swampland Criteria, where one quantum jump timescale exceeds the expansion rate of the universe during eternal inflation if $\eta>-\sqrt{3}$, which is marginally in tune with the alternative option provided by the refined de Sitter conjecture. But, as is it known, to explain the observations, eternal inflation should be followed by a slow-roll phase, in which case warm inflation turns out to be a a better option as has been pointed out in \cite{Kinney:2018kew}.

Let us now judge the option $ \eta \leq-c'$ with $\epsilon\ll1$ in the realm of warm inflation scenario. The scalar spectral tilt in warm inflation turns out to be \cite{Hall:2003zp}
\begin{eqnarray}
n_s-1=\frac1Q\left(-\frac94\epsilon+\frac32\eta-\frac94\beta\right),
\end{eqnarray}
where the slow-roll parameter $\beta$ is non-zero if the inflaton decay-width $\Gamma$ depends up on inflaton field $\phi$. With $\epsilon\ll1$, the above equation becomes
\begin{eqnarray}
n_s-1\approx\frac{3}{4Q}(2\eta-3\beta).
\end{eqnarray}
Since $\eta$ is negative and $|\eta|$ is of order unity as per the refined Swampland criterion, we see that $Q$ should be $\mathcal O(10)$ (irrespective of $\beta$ being identically zero or turns our to be of order unity) in order to keep warm inflation in accordance with observed scalar spectral tilt. Hence, we note that both the de Sitter options can be accommodated in warm inflation scenario, though the first option $\epsilon>c^2/2$ turns out to be more preferable as it doesn't demand $Q$ to be of $\mathcal O(10)$ \cite{Das:2018rpg}.

{\it In conclusion,} we point out that, despite the recent folklore that the single-field models are in tension with the String Swampland Criterion \cite{Agrawal:2018own, Achucarro:2018vey, Garg:2018reu, Dias:2018ngv, Kehagias:2018uem, Kinney:2018nny}, there are at least two single-field scenarios, such as non-Bunch-Davies initial condition for tensor modes \cite{Ashoorioon:2018sqb} and Warm inflation \cite{Berera:1995ie, Bartrum:2013oka}, which evade the apparent discrepancies to make themselves compatible with those criteria (even after the proposed refined version of the criteria \cite{Ooguri:2018wrx}). Warm inflation, in particular, turns out to be more interesting as such a scenario will survive even if the constant $c$ and $c'$ appearing in Swampland Criterion II turns out to be of $\mathcal{O}(1)$ in future, which is not the case for any other single-field slow-roll model. All in all, single-field scenario can potentially pass the litmus test of  Swampland Criteria with flying colours, provided we warm it up a bit. 

{\it Acknowledgements}:
 The work of S.D. is supported by Department of Science and Technology, 
 Government of India under the Grant Agreement number IFA13-PH-77 (INSPIRE Faculty Award). The author would like to thank Arjun Bagchi for useful discussions.

\label{Bibliography}
\bibliography{sland}

\end{document}